\documentstyle[aps,multicol]{revtex}
\newcommand{\be}{\begin{equation}}
\newcommand{\ee}{\end{equation}}
\newcommand{\ba}{\begin{eqnarray}}
\newcommand{\ea}{\end{eqnarray}}
\newcommand{\bas}{\begin{eqnarray*}}
\newcommand{\eas}{\end{eqnarray*}}
\begin{document}
%%%%%%%%%%%%%%%%%%%%%%%%%%%%%%%%%%%%%%%%%%%%%%%%%%%%%%%%%%%%%%%%%%%%%%%%%
\draft
%%%%%%%%%%%%%%%%%%%%%%%%%%%%%%%%%%%%%%%%%%
\title{
Exact dynamical structure factor of
the degenerate Haldane-Shastry model
}

\bigskip
\author{Takashi Yamamoto$^1$, Yasuhiro Saiga$^2$,
Mitsuhiro Arikawa$^3$, and Yoshio Kuramoto$^3$ }

\bigskip
\address{$^1$Max-Planck-Institut f\"{u}r Physik komplexer Systeme,
N\"{o}thnizer Str. 38, D-01187 Dresden, Germany\\
$^2$Institute for Solid State Physics, University of Tokyo,
Roppongi 7-22-1, Tokyo 106-8666, Japan\\
$^3$Department of Physics, Tohoku University,
Sendai 980-8578, Japan}
%%%%%%%%%%%%%%%%%%%%%%%%%%%%%%%%%%%%%%%%%%%%%%%%%%%%%%%%%%%%%%%%%%%%%%%%%
\date{\today}
%%%%%%%%%%%%%%%%%%%%%%%%%%%%%%%%%%%%%%%%%%%%%%%%%%%%%%%%%%%%%%%%%%%%%%%%%
\maketitle
%%%%%%%%%%%%%%%%%%%%%%%%%%%%%%%%%%%%%%%%%%%%%%%%%%%%%%%%%%%%%%%%%%%%%%%%%

\bigskip
\begin{abstract}
The dynamical structure factor $S(q,\omega)$ of the $K$-component
($K = 2,3,4$)
spin chain with
the $1/r^2$ exchange is derived exactly at zero temperature
for arbitrary size of the system.
The result is interpreted in terms of a free quasi-particle picture
which is generalization of the spinon picture in the $SU(2)$ case;
the excited states consist of $K$ quasi-particles each of
which is characterized by a set of $K-1$ quantum numbers.
Divergent singularities of $S(q,\omega)$ at the spectral edges are
derived analytically.
The analytic result is checked numerically for finite systems.
\end{abstract}
%%%%%%%%%%%%%%%%%%%%%%%%%%%%%%%%%%%%%%%%%%%%%%%%%%%%%%%%%%%%%%%%%%%%%%%%%
%%%%%%%%%%%%%%%%%%%%%%%%%%%%%%%%%%%%%%%%%%%%%%%%%%%%%%%%%%%%%%%%%%%%%%%%%

%%%%%%%%%%%%%%%%%%%%%%%%%%%%%%%%%%%%%%%%%%%%%%%%%%%%%%%%%%%%%%%%%%%%%%%%%
\begin{multicols}{2}
%%%%%%%%%%%%%%%%%%%%%%%%%%%%%%%%%%%%%%%%%%%%%%%%%%%%%%%%%%%%%%%%%%%%%%%%%

Recently much interest has been focused on magnetic systems with orbital
degeneracy
\cite{LMSZ,PSK,YSU,FMT,MFDT,AGLN}.
In the case of two-fold orbital degeneracy, the total degeneracy per
site becomes $4 \ (=2\times 2)$, and
the simplest model to realize this situation in one dimension is the
spin chain with $SU(4)$ symmetry.
The static property of the $SU(4)$ spin chain has been studied
mainly by numerical methods.
It has been reported that the spin correlation has a period of four unit
cells, and that the asymptotic decay
has a power-law exponent different from unity \cite{YSU,FMT}.
Such exponent has also been derived by use of conformal field theory
\cite{Affleck}.
In view of this situation, one can naturally ask how the dynamical
property depends on the number of
internal degrees of freedom.
Experimental investigations of
orbitally degenerate quasi-one-dimensional magnetic compounds
such as NaV$_2$O$_5$ \cite{Isobe}
are being performed with increasing accuracy.
Hence it is useful to clarify the difference from systems without
orbital degeneracy not only for static properties,
but also for dynamic ones.

In this Letter, we derive exact analytic formula for the dynamical
structure factor $S(q,\omega)$ of the
$SU(K)$ spin chain at zero temperature,
and provide intuitive interpretation of the result in terms of
quasi-particles obeying fractional statistics.
We take the exchange interaction $J_{ij}$ decaying as inverse-square of
the distance:
$J_{ij}=J[(N/\pi)\sin\pi(i-j)/N]^{-2}$ where $N$ is the number of
lattice sites with unit spacing and $J>0$.
The model is given by \cite{Kawakami,HaHaldane}
\be
\label{HS}
H_{\mbox{{\scriptsize HS}}}
=\frac{1}{2}\sum_{1\leq i<j\leq N}J_{ij}P_{ij},
\ee
where $P_{ij}$ is the exchange (or permutation) operator.
It can be written in the form:
\be
P_{ij}=\sum_{\delta,\gamma=1}^{K}X_i^{\delta\gamma}X_j^{\gamma\delta},
\ee
where $X_i^{\gamma\delta}$ changes the spin state $\delta$ to $\gamma$
at site $i$.
In the particular case of $SU(2)$,  $P_{ij}$ is reduced to the spin
exchange $2\vec S_i\cdot\vec S_j +1/2$.
This model is a generalization of the Haldane-Shastry (HS) model
\cite{HS1,HS2}
for the $SU(2)$ chain, and hence is called
the $SU(K)$ HS model in the following.

In the original HS model, the spinons form an ideal spin $1/2$ ``semion"
gas \cite{Haldane}
obeying  the fractional exclusion statistics \cite{fes}.
The dynamical structure factor $S(q,\omega)$ of the $SU(2)$ HS model
has a remarkably simple structure in terms of the spinon
picture:
only two spinons contribute to $S(q,\omega)$ \cite{HZ}.
Since the semionic statistics is applicable only to the case of
$SU(2)$,
one has to take more general fractional statistics in order to apply a
quasi-particle description.

To derive the exact formula for the dynamical structure factor,
we use the $U(K)$ spin Calogero-Sutherland (CS) model \cite{HaHaldane}
as an auxiliary.
The Hamiltonian of the $U(K)$ spin CS model is given by
\ba
\label{spinCS}
&&
H_{\mbox{{\scriptsize spinCS}}}=
\nonumber
\\
&&
-\frac{1}{2}\sum_{i=1}^N\frac{\partial^2}{\partial x_i^2}
+
\left(\frac{\pi}{L}\right)^2\sum_{1\leq i<j\leq N}
\frac{\beta(\beta+P_{ij})}{\sin^2\frac{\pi}{L}(x_i-x_j)},
\ea
where $\beta > 0$ is the coupling parameter and $L$ is the size of
system.
This continuous model is more tractable than the $SU(K)$ HS model,
because the eigenfunctions of the model have been explicitly
constructed \cite{TU,Uglov}.
We take the strong coupling limit $\beta\rightarrow\infty$ of the
$U(K)$ spin CS model. Then particles
crystallize with the lattice parameter $L/N$ which is taken as the unit
of length. Then we are left with the
center of mass motion, the lattice vibration and the dynamics of the
internal degrees of freedom which is
called the ``color''.
The color dynamics is equivalent to the dynamics of the $SU(K)$ HS
model.
The freezing trick described above was firstly introduced by
Polychronakos \cite{Polychronakos}, and
has been applied to thermodynamics of
lattice models
\cite{SS,KK1,KK2}.
The present Letter is the first application of the freezing trick to
dynamical quantities.

In the $U(2)$ spin CS model, Uglov has derived the exact formula
of the dynamical spin-density correlation function with a finite number
of particles \cite{Uglov}.
We shall first extend his result to the case of $K\geq 3$, and
then take the strong coupling limit.
In doing so we have to make correspondence between physical quantities
defined in the continuum and
discrete spaces.
Let us define the following
operator in the continuum space:
\be
X_q^{\gamma\delta}=\frac{1}{\sqrt L}\sum_{j=1}^N
X_{j}^{\gamma\delta}e^{-iqx_j},
\label{X_q}
\ee
where the momentum $q$ takes values $2\pi n/L$ with $n$ an arbitrary
integer.
We first derive the dynamical structure factor in the continuum model
defined by
\be
\label{def-dssf}
S^{(\gamma\delta)}(q,\omega;\beta)
=
\sum_{\alpha}|\langle\alpha|X_q^{\gamma\delta}|0\rangle|^2
\delta(\omega-E_\alpha+E_0),
\ee
where $\{|\alpha\rangle\}$ is the normalized complete basis of the
system with eigenvalues $\{E_\alpha\}$,
and $|0\rangle$ is the groundstate.
We assume that $N$ is an integer multiple of $K$ so that the groundstate
is a nondegenerate singlet.

In the strong coupling limit the
coordinate $x_j$ in Eq. (\ref{X_q}) is written as
$x_j=R_j+u_j$ where $R_j =j $ is a lattice point, and $u_j$
describes the lattice vibration.
Except for the uniform motion of the lattice
described by $u_j= const$,
we may regard $u_j$ as a small quantity.
In fact the density response can be shown to be smaller than the spin
response by ${\cal O}(\beta^{-1})$.
Then the dynamical structure factor of the $SU(K)$ HS model is given
simply by the strong coupling limit of
Eq. (\ref{def-dssf}) provided that one restricts $q$ in the range of the
first Brillouin zone: $|q| \le \pi$.

The dynamical structure factor (\ref{def-dssf})
can be derived in a manner analogous to
the case of $K=2$ \cite{Uglov}.
However, the following observations are necessary for generalization.
First, each excited state relevant to Eq. (\ref{def-dssf})
transforms as one of the weight vectors for
the adjoint representation of $SU(K)$.
This observation allows us to find the selection rule for the $SU(K)$ spin.
In the $SU(2)$ case, this selection rule is reduced to
the simple fact that excited states relevant to Eq. (\ref{def-dssf})
are spin-triplet states.
Second, in order to derive the matrix element in Eq. (\ref{def-dssf})
we find a convenient set of operators given by
%%change
\begin{equation}
J_a =
\frac{1}{\sqrt{L}}
\sum_q
\left(
\sum_{b=1}^a X_q^{b,K-a+b}
+
\sum_{b=1}^{K-a} X_q^{a+b,b}
\right)
\end{equation}
with $a=1,\cdots,K-1$.
More details of calculation will be presented elsewhere.

In order to give the formula of $S^{(\gamma\delta)}(q,\omega)$,
we fix some notations for partitions\cite{Uglov}.
Let $\Lambda^{(K)}_N$
be the set of all partitions whose length are less than $N+1$
and the largest entry is less than $K+1$.
Namely we have
$\Lambda^{(K)}_N
=\{\lambda=(\lambda_1,\lambda_2,\cdots,\lambda_N)
|\,K\geq\lambda_1\geq\lambda_2\geq\cdots\geq\lambda_N\geq 0\}$.
For a partition $\lambda$, we define subsets by
$C_K(\lambda)
=\{(i,j)\in\lambda\,|\,j-i\equiv 0\ \mbox{mod}\, K\}$
and
$H_K(\lambda)
=\{(i,j)\in\lambda\,|\,\lambda_i+\lambda'_j-i-j+1
\equiv 0\ \mbox{mod}\, K\}$.
Next we define the concept {\it type} of a partition.
The {\it type} of given partition $\lambda$ contains sufficient
information for
determining the sets $C_K(\lambda)$ and $H_K(\lambda)$ explicitly.
We introduce a reductive transformation $\tau$
on the set of all partitions as follows \cite{YA}:\\
(i) If there exist $K$ rows or $K$ columns which have same
number of boxes in a partition,  remove those rows or columns;\\
(ii) Apply the reduction (i) repeatedly until the newly generated
partition is no longer reducible.\\
We then determine a subset ${\cal A}^{(K)}_N$ of $\Lambda^{(K)}_N$
as the image of $\tau$, i.e., ${\cal A}^{(K)}_N=\tau(\Lambda^{(K)}_N)$.
For any partition $\lambda\in\Lambda^{(K)}_N$,
we say that $\nu$ is of the {\it type} $\lambda$ if
$\nu=\tau(\lambda)$.
The total number of {\it types} increases from 3 in the case of $SU(2)$
to 25 in $SU(3)$, and
to 252 in $SU(4)$.
For a box $s=(i,j)\in\lambda$, the numbers
$l_\lambda(s)=\lambda'_j-i$ and $l_\lambda'(s)=i-1$
are called the leg-length and coleg-length, respectively.
For any subset $\nu\subset\lambda$, the order $|\nu|$ is defined as
the number of boxes in $\nu$.

Now we give the exact
formula for the dynamical structure factor
$S^{(\gamma\delta)}(q,\omega)$
for a finite size of the system with $\gamma\ne\delta$.
We obtain
\ba
\label{finite-dssf}
S^{(\gamma\delta)}(q,\omega)
&=&
\sum_\lambda{}^{{\normalsize '}}
|F_\lambda^{(K)}|^2
\delta(\omega-E_\lambda),
\ea
where the primed summation is restricted so as to satisfy the momentum
conservation
$q=2\pi |C_K(\lambda)|/N$, and the color selection rule for
$\lambda \in \Lambda_N^{(K)}$.
The latter is conveniently implemented by introducing
a subset ${\cal A}_N^{(K;\gamma\delta)}$ of ${\cal A}_N^{(K)}$,
and decomposes the summation over $\lambda$ by each {\it type}
$\nu = \tau (\lambda)$ such that $\nu\in {\cal A}_N^{(K;\gamma\delta)}$.
For example, in the case of $K=3$, we have
${\cal A}_N^{(3;21)}=\{(2,1,1),(3,3,1),(3,3,2,2)\}$ and
${\cal A}_N^{(3;13)}=\{(1),(2,2),(3,2,1,1)\}$.

In Eq. (\ref{finite-dssf}) the excitation energy is given by
\ba
E_\lambda
=\frac{J}{4}\left(\frac{2\pi}{N}\right)^2
\Big[(N-1)|C_K(\lambda)|-2\sum_{s\in C_K(\lambda)}l_\lambda'(s)\Big],
\ea
and the squared form factor by
\ba
|F_\lambda^{(K)}|^2 &=&  \frac{1}{N}
\frac{\prod_{s\in C_K(\lambda)\setminus\{(1,1)\}}l_\lambda'(s)^2}
     {\prod_{s\in H_K(\lambda)}l_\lambda(s)(l_\lambda(s)+1)} \nonumber
\\
&& \times \prod_{s\in
C_K(\lambda)}\frac{N-l_\lambda'(s)}{N-l_\lambda'(s)-1}.
\label{finite-ff}
\ea

Since we consider the case of zero external magnetic fields,
the $SU(K)$ symmetry demands
that $S^{(\gamma\delta)}(q,\omega)$ is actually independent of
$(\gamma,\delta)$ as long as $\gamma\ne\delta$.
We can prove this fact using the expressions
(\ref{finite-dssf})-(\ref{finite-ff}).
In the particular case of $K=2$, our formulae
(\ref{finite-dssf})-(\ref{finite-ff})
generalize the known one \cite{HZ} to arbitrary size of the system.
We have checked the validity of Eqs.
(\ref{finite-dssf})-(\ref{finite-ff}) with $K=2$ and 3
by comparing with the numerical result for $N\leq 24$ and $N\leq 15$,
respectively.
The numerical result
is obtained via exact diagonalization and the recursion
method.
The agreement is excellent in both cases of $K=2$ and 3.
In Fig. \ref{fig1}, we present the result for $K=3$ and $N=15$.

We now consider the quasi-particle interpretation of the color
selection rule.
For labelling the excited states relevant to
$S^{(\gamma\delta)}(q,\omega)$,
it is more convenient to use the conjugate partition
$\lambda'=(\lambda'_1,\cdots,\lambda'_K)\in\Lambda_K^{(N)}$
instead of $\lambda\in\Lambda_N^{(K)}$.
Each $\lambda'_i$ has the information on the momentum and $SU(K)$ spin
of a quasi-particle.
We call this quasi-particle a spinon following the $SU(2)$ case.
The spinon is considered to be an object possessing the $SU(K)$ spin.
Here the $SU(K)$ spin means
the $K-1$ eigenvalues $(s_{1},\cdots,s_{K-1})$
of a set of operators $(h^{1},\cdots,h^{K-1})$ where
$h^\gamma$ is defined by
$h^\gamma
=\sum_{i=1}^N(X_i^{\gamma\gamma}-X_i^{\gamma+1 \gamma+1})/2$
for $\gamma=1,\cdots,K-1$.
For the $SU(2)$ case, this definition gives
the $z$-component of spin.
The $SU(K)$ spin of the spinon with $\lambda'_i$ is specified by
a certain condition on the pair $(i,\lambda'_i)$.
Since the condition for general $K$
is rather complicated,
we give an example in the case of $K=3$.
The $SU(3)$ spin for $\lambda'_i$ is assigned as follows:
It is
$(0,1/2)$ if
$(i,\lambda'_i)\equiv (0,0), (1,1), (2,2) \mbox{ mod } 3$;
$(1/2,-1/2)$ if
$(i,\lambda'_i)\equiv (0,1), (1,2), (2,0) \mbox{ mod } 3$;
and
$(-1/2,0)$ if
$(i,\lambda'_i)\equiv (0,2), (1,0), (2,1) \mbox{ mod } 3$.
It is important to note that
a spinon transforms as a weight vector
of the fundamental representation $\bar{K}$
of $SU(K)$.

{}From the formulae (\ref{finite-dssf})-(\ref{finite-ff}),
we can conclude that relevant excited states for the $SU(K)$ HS model
consist of $K$ spinons.
Moreover the conditions on the {\it type} of excited states
lead to an important consequence:
$K$-spinon excited states have $K-1$ different $SU(K)$ spins.
That is, the excited states contain $K-1$ species of quasi-particles.
For instance, in the case of $K=3$
excited states relevant to $S^{(13)}(q,\omega)$
consist of three spinons with $SU(3)$ spins $(0,1/2)$, $(0,1/2)$
and $(1/2,-1/2)$.
For the $SU(2)$ case, we recover the well-known fact
that only two spinons with the same spin contribute
to $S(q,\omega)$ \cite{HZ}.

The $K$-spinon excitation belongs to
the tensor representation $\bar{K}^{\otimes K}$ of $SU(K)$.
This representation
contains the adjoint representation as
an irreducible component. From
the condition stated above for the $K$-spinon excitation,
we see that the $K$-spinon excitation transforms
as one of the weight vectors for the adjoint representation.
This is consistent with the condition for the $SU(K)$ spins
of the excited states which are relevant to
$S^{(\gamma\delta)}(q,\omega)$.

Now we present the thermodynamic limit of
the formulae (\ref{finite-dssf})-(\ref{finite-ff}).
Performing a procedure similar to that in Refs. \cite{Ha,LPS},
for $K=2,3$ and $4$,
we obtain the final result as follows:
\ba
S(q,\omega)
&=&
A_K \sum_{1\leq a<b\leq K}
\prod_{i=1}^K\int_{-1}^{1}dk_i
|F_{ab}^{(K)}(k)|^2 \nonumber\\
&& \times \delta(q-\pi-p(k))\delta(\omega-\epsilon(k)),
\label{thermo-dssf}
\ea
where $\epsilon(k)=[\pi v_s/(2K)]\sum_{i=1}^K(1-k_i^2)$
with $v_s=J\pi/2$, and
$p(k)=(\pi/K)\sum_{i=1}^K k_i$.
In the above formula,
$A_K$ is a normalization constant given by
\ba
\label{thermo-const}
A_K
&=&
\frac{2^{K}\pi}{K^3(K-1)}
\prod_{j=1}^K\frac{\Gamma((K-1)/K)}{\Gamma(j/K)^2},
\ea
and the form factor is given by
\ba
\label{thermo-ff}
&&
F_{ab}^{(K)}(k)
=
\nonumber
\\
&&
\frac{|k_a-k_b|^{g_K}
      \prod_{1\leq i<j\leq K,(i,j)\ne(a,b)}|k_i-k_j|^{g'_K}}
     {\prod_{i=1}^K(1-k_i^2)^{(1-g_K)/2}}
\ea
with $g_K=(K-1)/K$ and $g'_K=-1/K$.
Since the formula (\ref{thermo-dssf}) for
$S^{(\gamma\delta)}(q,\omega)$
does not depend on the pair $(\gamma,\delta)$ with $\gamma\ne\delta$, we
have omitted the superscript.
Unfortunately our exact result is inconsistent with a conjecture
proposed several years ago \cite{conjecture}.

For $K=2$, the formula reproduces the result
of Haldane-Zirnbauer \cite{HZ}
which was obtained by a completely different method.
Notice that the second product in the numerator
of Eq. (\ref{thermo-ff}) is absent in the $SU(2)$ case.

We can derive static structure factor $S(q)$ by integrating over
$\omega$ in Eq. (\ref{thermo-dssf}).
In the low energy limit we recover the results of Ref. \cite{Kawakami}
which are obtained by conformal field theory.
By analyzing the form factor for $q \sim 2k_F$ with $k_F=\pi/K$, we can
show that it has the asymptotic form
\be
S(q)\sim a_1|q-2k_F|^{\alpha_1-1}
\ee
with the exponent $\alpha_1= 2-2/K$ and a non-universal constant $a_1$.
In the real space the spin (or color) correlation decays as
$b_1 \cos(2k_Fx)|x|^{-\alpha_1}$ with
a certain coefficient $b_1$.
Similar analysis shows that there are also weaker singularities around
$q =
2lk_F$
for $l=2,\cdots,K-1$
with exponents $\alpha_l=2l(1-l/K)$.
These $K-1$ singularities correspond to $K-1$ gapless bosonic modes
\cite{Affleck,Kawakami}

The spinon interpretation of the formula (\ref{thermo-dssf})
goes as follows.
As in the case of finite systems,
the excited states for $S(q,\omega)$ in the thermodynamic limit
consist of $K$ spinons with $K-1$ different $SU(K)$ spins.
This fact means, as in the low energy limit,
the dynamics of the $SU(K)$ HS model can be described by $K-1$ species
of quasi-particles.  This simple structure reflects
the Yangian symmetry of the $SU(K)$ HS model \cite{HHTBP}.
In the form factor (\ref{thermo-ff}),
the factor $|k_a-k_b|^{g_K}$
represents the statistical interactions of spinons with the same $SU(K)$
spin, while the factor $|k_i-k_j|^{g'_K}$
represents those of spinons with different $SU(K)$ spins.
We refer to Ref. \cite{KYA}
for more detailed explanation of statistical interactions.
It will be interesting to consider the relation between our results
and the exclusion statistics in conformal field theory
discussed in Refs. \cite{BS,Suzuki}.

The support of $S(q,\omega)$ represents the region
in the momentum-frequency plane where $S(q,\omega)$ takes the non-zero
value.
We see that the support of $S(q,\omega)$ as determined from the formula
(\ref{thermo-dssf}) is compact,
i.e., there is no intensity outside of the finite area.
In the $SU(3)$ case, for example, the support is determined as:\\
$\omega\leq[v_s/(2\pi)]q(2\pi-q)
\equiv
\epsilon^{(\mbox{{\tiny U}})}(q)$
for $0\leq q\leq 2\pi$;\\
$\omega\geq[3v_s/(2\pi)]q(2\pi/3-q)$ for $0\leq q\leq 2\pi/3$;\\
$\omega\geq[3v_s/(2\pi)](q-2\pi/3)(4\pi/3-q)$ for $2\pi/3\leq q\leq
4\pi/3$;\\
$\omega\geq[3v_s/(2\pi)](q-4\pi/3)(2\pi-q)$ for $4\pi/3\leq q\leq
2\pi$.\\
For a general value of $K$, there are $K$ lower boundaries as given by
$$
\epsilon_j^{(\mbox{{\tiny L}})}(q)  \equiv
[Kv_s/(2\pi)](q-2\pi j/K) [2\pi (j+1)/K-q],
$$
for $2\pi j/K\leq q\leq 2\pi(j+1)/K$ with $j=0,1,\ldots ,K-1$.

The behavior of $S(q,\omega)$ near the boundaries of the support
is derived for general $K$ as follows.
We can show that there is a stepwise discontinuity at the upper boundary
$\omega=\epsilon^{(\mbox{{\tiny U}})}(q)$.
On the other hand, there are divergent singularities at the lower
boundaries
$\omega
= \epsilon_0^{(\mbox{{\tiny L}})}(q)$ and $\omega
= \epsilon_{K-1}^{(\mbox{{\tiny L}})}(q)$.
Here $S(q,\omega)$ diverges by the power law with the exponent $-1/K$.
At the other lower boundaries $\omega = \epsilon_j^{(\mbox{{\tiny L}})}(q)$
with $j\neq 0, K-1$,
$S(q,\omega)$ has threshold singularities but no divergence.
As in the $SU(2)$ case \cite{HZ,KMBFM},
we expect that the divergences at two of the lower boundaries
occur also in the $SU(K)$ Heisenberg model with
the nearest-neighbor exchange.

In conclusion,
we have derived the exact formulae
(\ref{finite-dssf}) and (\ref{thermo-dssf}) for $S(q,\omega)$ of
the $SU(K)$ HS model for arbitrary size of the system
at zero temperature.
Our exact result of $S(q,\omega)$ for $K \leq 4$ is likely to be valid
for larger $K$ as well.
We have also clarified the quasi-particle picture
of the spin dynamics.
The relevant excited states consist of $K$ spinons with $K-1$ different
$SU(K)$ spins.

T.Y. and Y.S. wish to acknowledge the support of the CREST from the
Japan Science and Technology Corporation.
T.Y. wishes to thank the support of the Visitor Program of the MPI-PKS.

%%%%%%%%%%%%%%%%%%%%%%%%%%%%%%%%%%%%%%%%%%%%%%%%%%%%%%%%%%%%%%%%%%%%%%%%%

%%%%%%%%%%%%%%%%%%%%%%%%%%%%%%%%%%%%%%%%%%%%%%%%%%%%%%%%%%%%%%%%%%%%%%%%%

%%%%%%%%%%%%%%%%%%%%%%%%%%%%%%%%%%%%%%%%%%%%%%%%%%%%%%%%%%%%%%%%%%%%%%%%%
%\end{multicols}
%%%%%%%%%%%%%%%%%%%%%%%%%%%%%%%%%%%%%%%%%%%%%%%%%%%%%%%%%%%%%%%%%%%%%%%%%

\begin{figure}
\caption[]{
Numerical result of the dynamical structure factor $S(q,\omega)$
in the case of $K=3$ and $N=15$.
The vertical and horizontal axis represent
the rescaled energy and momentum, respectively.
The intensity is proportional to the area of the circle.
The solid lines are the dispersion lines of the elementary excitations
in the thermodynamic limit.
The analytic results are in excellent agreement with numerical ones, and
are not distinguishable from the latter.}
\label{fig1}
\end{figure}

%%%%%%%%%%%%%%%%%%%%%%%%%%%%%%%%%%%%%%%%%%%%%%%%%%%%%%%%%%%%%%%%%%%%%%%%%
\end{multicols}
%%%%%%%%%%%%%%%%%%%%%%%%%%%%%%%%%%%%%%%%%%%%%%%%%%%%%%%%%%%%%%%%%%%%%%%%%

%%%%%%%%%%%%%%%%%%%%%%%%%%%%%%%%%%%%%%%%%%%%%%%%%%%%%%%%%%%%%%%%%%%%%%%%%

\begin{references}

%% orbital degeneracy

%%%
\bibitem{LMSZ}
Y. Q. Li, M. Ma, D. N. Shi and F. C. Zhang,
Phys. Rev. Lett. {\bf 81}, 3527 (1998).

%%%
\bibitem{PSK}
S. K. Pati, R. R. P. Singh and D. I. Khomskii,
Phys. Rev. Lett. {\bf 81}, 5406 (1998).

%%%
\bibitem{YSU}
Y. Yamashita, N. Shibata and K. Ueda,
Phys. Rev. B {\bf 58}, 9114 (1998).

%%%
\bibitem{FMT}
B. Frischmuth, F. Mila and M. Troyer,
Phys. Rev. Lett. {\bf 82}, 835 (1999).

\bibitem{MFDT}
F. Mila, B. Frischmuth, A. Deppeler and M. Troyer,
Phys. Rev. Lett. {\bf 82}, 3697 (1999).


%%%
\bibitem{AGLN}
P. Azaria, A. O. Gogolin, P. Lecheminant and A. A. Neresyan,
Phys. Rev. Lett. {\bf 83}, 624 (1999).


%%% Affleck
\bibitem{Affleck}
I. Affleck,
Nucl. Phys. B {\bf 265}, 409 (1986).

%%% NaV2O5
\bibitem{Isobe}
M. Isobe and Y. Ueda,
J. Phys. Soc. Jpn. {\bf 65}, 1178 (1996).

%% spin CS model and SU(K) HS model

\bibitem{Kawakami}
N. Kawakami,
Phys. Rev. B {\bf 46}, 1005 (1992).

\bibitem{HaHaldane}
Z. N. C. Ha and F. D. M. Haldane,
Phys. Rev. B {\bf 46}, 9359 (1992).


%% SU(2) HS model

\bibitem{HS1}
F. D. M. Haldane,
Phys. Rev. Lett. {\bf 60}, 635 (1988).

\bibitem{HS2}
B. S. Shastry,
Phys. Rev. Lett. {\bf 60}, 639 (1988).


%% semion gas
\bibitem{Haldane}
F. D. M. Haldane,
Phys. Rev. Lett. {\bf 66}, 1529 (1991).


%% fractional exclusion statistics
\bibitem{fes}
F. D. M. Haldane,
Phys. Rev. Lett. {\bf 67}, 937 (1991).


%% SU(2) HS dynamics
\bibitem{HZ}
F. D. M. Haldane and M. R. Zirnbauer,
Phys. Rev. Lett. {\bf 71}, 4005 (1993).


%% Takemura-Uglov
\bibitem{TU}
K. Takemura and D. Uglov,
J. Phys. A: Math. Gen. {\bf 30}, 3685 (1997).

%% Uglov
\bibitem{Uglov}
D. Uglov,
Commun. Math. Phys. {\bf 191}, 663 (1998).


%% freezing trick
\bibitem{Polychronakos}
A. P. Polychronakos,
Phys. Rev. Lett. {\bf 70}, 2329 (1993).


%% applications of the freezing trick
\bibitem{SS}
B. Sutherland and B. S. Shastry,
Phys. Rev. Lett. {\bf 71}, 5 (1993).


\bibitem{KK1}
Y. Kuramoto and Y. Kato,
J. Phys. Soc. Jpn. {\bf 64}, 4518 (1995).
\bibitem{KK2}
Y. Kato and Y. Kuramoto,
J. Phys. Soc. Jpn. {\bf 65}, 1622 (1996).


%%%
\bibitem{YA}
T. Yamamoto and M. Arikawa,
J. Phys. A: Math. Gen. {\bf 32}, 3341 (1999).

%%% using Jack
\bibitem{Ha}
Z. N. C. Ha,
Phys. Rev. Lett. {\bf 73}, 1574 (1995);
{\it ibid.} {\bf 74}, 620 (1995) (Errata); Nucl. Phys. B {\bf 435}, 604
(1995).
\bibitem{LPS}
F. Lesage, V. Pasquier and D. Serban,
Nucl. Phys. B {\bf 435}, 585 (1995).

%% conjecture
\bibitem{conjecture}
F. D. M. Haldane,
in
{\it Correlation Effects in Low-Dimensional Electron Systems},
edited by A. Okiji and N. Kawakami (Springer Verlag, Berlin, 1994).


%%% Yangian & SU(K) HS model
\bibitem{HHTBP}
F. D. M. Haldane, Z. N. C. Ha, J. C. Talstra, D. Bernard and V.
Pasquier,
Phys. Rev. Lett. {\bf 69}, 2021 (1992).


%%% KYA

\bibitem{KYA}
Y. Kato, T. Yamamoto and M. Arikawa,
J. Phys. Soc. Jpn. {\bf 66}, 1954 (1997).


%%% g-matrix

\bibitem{BS}
P. Bouwknegt and K. Schoutens,
Nucl. Phys. B {\bf 547}, 501 (1999).


\bibitem{Suzuki}
J. Suzuki,
J. Phys. A: Math. Gen. {\bf 31}, 6887 (1999).

\bibitem{KMBFM}
M. Karbach, G. M\"{u}ller, A. H. Bougourzi, A. Fledderjohann
and K.-H. M\"{u}tter,
Phys. Rev. B {\bf 55}, 12510 (1997).

\end{references}
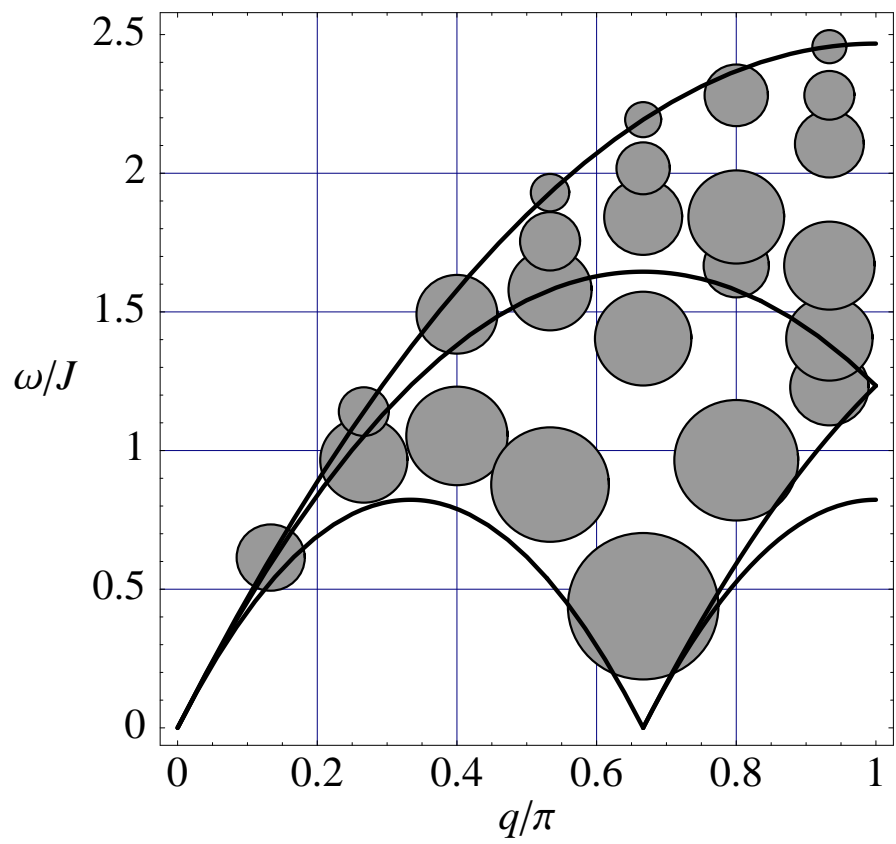
\end{document}